\documentclass[
twocolumn, 
superscriptaddress,
nofootinbib,
amsmath,amssymb,
longbibliography,
prx,
]{revtex4-2}

\usepackage[utf8]{inputenc}
\usepackage{CJKutf8}
\usepackage{graphicx}
\usepackage{subfigure}
\usepackage{color}
\usepackage[dvipsnames]{xcolor}
\usepackage{amsmath,amssymb,bm}
\usepackage{cases}
\usepackage{tikz}
  \usetikzlibrary{calc}
  \usetikzlibrary{arrows}
  \usetikzlibrary{decorations.markings}
\usepackage{afterpage}
\usepackage{ulem}
\usepackage{hyperref}

\expandafter\ifx\csname package@font\endcsname\relax\else
  \expandafter\expandafter
  \expandafter\usepackage
  \expandafter\expandafter
  \expandafter{\csname package@font\endcsname}%
\fi

\usepackage{dsfont}

\newcommand{\canetset}[1]{{\mathchoice {\hbox{$\sf\textstyle #1\kern-0.4em #1$}}
{\hbox{$\sf\textstyle #1\kern-0.4em #1$}}
{\hbox{$\sf\scriptstyle #1\kern-0.3em #1$}}
{\hbox{$\sf\scriptscriptstyle #1\kern-0.2em #1$}}}}

\begin{document}

\title{The emergence of Newtonian mechanics from the inhomogeneity of an ensemble}

\author{Hong Yuan (\begin{CJK}{UTF8}{gbsn}袁红\end{CJK})}
% \email{yuanhong18@gscaep.ac.cn}
\affiliation{Graduate School of CAEP, Beijing 100193, China}
% \affiliation{}

\author{C.P. Sun (\begin{CJK}{UTF8}{gbsn}孙昌璞\end{CJK})}
\email{suncp@gscaep.ac.cn}
\affiliation{Graduate School of CAEP, Beijing 100193, China}

\begin{abstract}
To address the observation of Max Born (M. Born 1969) that the Newton's second law can emerge from a purely statistical perspective, we derive the evolution equation about the statistical distribution for dilute gas  based solely on statistical principles, without invoking Newtonian mechanics, and then obtain the equations of motion for individual particles. Newton's second law for a single particle naturally emerges when the distribution reaches equilibrium. We demonstrate that the magnitude of an external force, traditionally measured by particle acceleration, can be understood as a measure of distribution inhomogeneity. We further show that the entropic force ( utilised in current gravity studies) is equivalent to the statistical force and under non-equilibrium conditions, a deviation arises between the entropic force and the Newtonian force. This framework offers a novel perspective distinct from classical Newtonian mechanics and broadens the potential scope of its application.

%We explore how Newton's second law emerges from the statistical perspective, which is suggested by Max Born\cite{BornMax1969Pimg}. We derive the evolution of the distribution for a dilute gas system purely from statistical principles without using Newtonian mechanics. Consequently, the equation of motion of each particle are in turn obtained and the Newton's second law for one particle naturally emerges when the distribution reaches equilibrium. It is shown that the magnitude of the external force, classically measured by the acceleration of a particle, can be regarded as a measure of the heterogeneity of the distribution. Our theory provides a new and innovative perspective different from the traditional Newtonian mechanics framework, and thus has a wide range of application scenarios.

\end{abstract}

% \keywords{Self-organised criticality,
% Universality,
% Correlation functions,
% Finite-size scaling}

\maketitle

%\tableofcontents

\section{\label{sec level1}INTRODUCTION}

The concept of force, while central to classical mechanics, has remained somewhat abstract since its introduction in Newton's second law. Newton defined force as the product of mass and acceleration, establishing a deterministic relationship between an object's motion and the forces acting upon it. However, while this law provides a method to calculate force, it does not offer a deeper understanding of what the force *is*\cite{wilczek2004whence1,wilczek2004whence}. Newton's second law describes how force affects the motion of a body, but it leaves the nature of force itself largely unexplored, treating it as an external, predefined entity without addressing its origin or underlying characteristics. This longstanding question has inspired various perspectives throughout the history of physics. For instance, Max Born demonstrated, through the analysis of $\alpha$-particle scattering and the barometer formula, that forces can be evaluated statistically—by counting and distribution analysis—rather than solely by measuring acceleration, as prescribed by Newton’s law. Born suggested that Newtonian force could be redefined in statistical terms: just as uniform rectilinear motion in classical mechanics implies the absence of external forces, a uniform distribution of particles may indicate a force-free state for each particle. Here, the magnitude of force, traditionally gauged by particle acceleration, would instead be measured by the degree of inhomogeneity in the particle distribution \cite{BornMax1969Pimg}. Similarly, Einstein’s investigation of Brownian motion \cite{Einstein_Browian_motion} established a connection between force and the density gradient, directly linking force to the ununiformity of a system's distribution.

From a broader perspective, while Newtonian mechanics has been remarkably successful in describing classical systems, revisiting these foundational concepts remains valuable. Unconventional and innovative approaches can offer fresh insights and potentially lead to theoretical breakthroughs, as demonstrated by the advent of quantum mechanics and relativity. For example, the determinism inherent in classical mechanics, though widely accepted, does not rule out the possibility of a probabilistic description even at the classical level. Phenomena such as Brownian motion suggest that probabilistic frameworks may be more suitable in certain contexts \cite{wang1945theory,uhlenbeck1930theory,parisi1981perturbation}. Recently, Bing Miao, Hong Qian, and Yong-Shi Wu proposed that Newtonian deterministic causality can emerge from underlying stochastic dynamics\cite{miao2024emergence}. Similarly, Verlinde’s work on entropic forces \cite{Verlinde}, which derives gravity from changes in information associated with the positions of matter, highlights how thermodynamics and information theory can offer new perspectives. His approach has significantly influenced cosmological research on gravity and black holes\cite{bahamonde2023teleparallel,famaey2012modified,verlinde2017emergent,amelino2013quantum}, underscoring the compatibility of probabilistic descriptions with classical physical reality.

In this paper, inspired by Born, Einstein and even Verlinde's perspectives, we propose a statistical redefinition of force as a measure of the ununiformity of a system’s distribution in phase space. Rather than assuming that particles move deterministically according to Newton’s second law, we begin from the premise that particles exist in interaction with their environment, and their motion is inherently uncertain. To describe the state of the particles, we use distribution functions that account for their position and velocity, acknowledging the stochastic nature of their dynamics. By considering the evolution of these distributions over time, we derive a set of equations that govern the changes in the system's state without invoking Newton’s laws. The Fokker-Planck equation, central to this approach, allows us to track the time evolution of the probability distribution of particle states in phase space. What emerges from this statistical framework is a natural, emergent appearance of Newton's second law as a limiting case when the system reaches equilibrium. At this point, this approach offers a new perspective on force, framing it not as a fundamental entity in its own right, but as a statistical property of the system's distribution in phase space. By moving beyond the deterministic framework of Newtonian mechanics, this theory opens the door to a broader range of applications, particularly in systems where uncertainty and stochastic interactions dominate, from thermodynamic systems to complex, non-equilibrium environments.

The paper is structured as follows. In Section \ref{The Evolution equation of the distribution for a dilute gas system}, we derive the evolution equation for the distribution of a dilute gas system based purely on statistical principles, without invoking Newtonian mechanics. From this, we obtain the equations of motion for individual particles. Section \ref{Statistical force and its generalization} defines the concept of force from a statistical perspective, demonstrating how Newton’s second law naturally emerges when the system reaches equilibrium. We further extend the concept of statistical force by introducing a dissipative force that represents the average influence of the environment on the system. To provide a concrete understanding of statistical force, two illustrative examples are presented in Section \ref{Two illustrations of the statistical force}. Finally, Section \ref{conclusion} summarizes the main findings.

\section{Revisit Derivation of Time-dependent Distribution for a Dilute gas system}
\label{The Evolution equation of the distribution for a dilute gas system}

In this section, we show how the equations of motion can be derived purely from statistical principles, bypassing the need for traditional mechanical definitions. In specific, we derive the evolution equation of the distribution for a dilute gas system based solely on statistical principles, without invoking Newtonian mechanics, and then obtain the equations of motion for individual particles.

We consider a dilute system of particles subject to an external potential $V(x)$, where the particle number density $n = N/V$ is sufficiently small. In this limit, interactions between particles can be neglected, and the particles can be treated as statistically independent, similar to an ideal gas. The particle number density at time $t$ is denoted by $n(x,t)$. For simplicity, we restrict our analysis to one spatial dimension.

Let $f(x,v,t)$ denote the single-particle distribution function in phase space (the $\mu$-space), where $d\omega = dx\, dv$ represents a phase space volume element and $\Omega$ is the total phase space volume. The quantity $f(x,v,t)\, d\omega$ gives the probability of finding a particle in the phase space volume $d\omega$ near $(x,v)$ at time $t$. For a system of $N$ statistically independent and identical particles, each particle's distribution function is the same in its respective phase space. Therefore, we can discuss the problem in single-particle phase space, where $Nf(x,v,t)\, d\omega$ represents the number of particles in the phase space volume $d\omega$ near $(x,v)$ at time $t$.

The evolution of the distribution function is driven by two main factors, i.e., changes due to the external potential and modifications due to collisions between particles. In the following, we derive the evolution equations for both cases separately.

%In this section, we redefine the concept of "force" within the framework of a canonical ensemble. We show how the equations of motion can be derived purely from statistical principles, bypassing the need for traditional mechanical definitions. As a result, when the ensemble reaches equilibrium state, the statistical force is equal to the Newtonian force.

%\subsection{Evolution equation of the distribution due to collisions}
%\label{subsection-Evolution of the Distribution Function and Equation of Motion Due to Collisions}

Firstly, we revisit how to derive the evolution equation of the distribution from purely probability method. Specifically, we focus on the case of small displacements in each micro-step. Let $f(x,v,t)$ denote the distribution function at time $t$. Over the time interval $\Delta t$, the probability of a change in position by $\Delta x$ due to collisions is $\varphi(\Delta x)$, and the probability of a change in velocity by $\Delta v$ is $\psi(\Delta v)$. Generally, $\varphi(\Delta x)$ and $\psi(\Delta v)$ depend on the pre-collision state, such that $\varphi(\Delta x) = \varphi(x,v,\Delta x)$ and $\psi(\Delta v) = \psi(x,v,\Delta v)$. Since collisions are assumed to be symmetric, we have $\varphi(\Delta x) = \varphi(-\Delta x)$, indicating that the particle has an equal probability of moving left or right.

The $k$-th moments of the position and velocity changes over the interval $\Delta t$ are defined as 
\begin{align}
\overline{\Delta x^{k}} &\equiv \int \Delta x^{k} \varphi(\Delta x) d(\Delta x), \\
\overline{\Delta v^{k}} &\equiv \int \Delta v^{k} \psi(\Delta v) d(\Delta v),
\end{align}
which represent the $k$-th order moments of the displacements in position and velocity. For small displacements, i.e., as $\Delta t \to 0$, the third and higher moments are higher-order infinitesimals in $\Delta t$. Note that $\overline{\Delta x \Delta v}= \mathcal{O}(\Delta t^{2})$ and $\overline{\Delta x} = 0$, so the resulting evolution of the distribution function is given by the Fokker-Planck equation\cite{VanKampen,Risken} 
\begin{align}
\frac{\partial f}{\partial t} &= -\frac{\partial}{\partial v} \left[ f A(x,v) \right] + \frac{\partial^{2}}{\partial v^{2}} \left[ f B(x,v) \right],\label{eq coll}
\end{align}
where
\begin{equation}
    A(x,v) \equiv \lim_{\Delta t \to 0} \frac{\overline{\Delta v}}{\Delta t}, \quad B(x,v) \equiv \lim_{\Delta t \to 0} \frac{1}{2} \frac{\overline{\Delta v^{2}}}{\Delta t}.
\end{equation}
At equilibrium, the equilibrium condition, $\partial f/\partial t = 0$, yields
\begin{equation}
f_{eq}(x,v) = \frac{A}{B(x,v)} \exp \left[ \int_{0}^{v} \frac{A(x,v')}{B(x,v')} dv' \right], \label{eq f_eq}
\end{equation}
where $A$ is a normalized constant. In the absence of external forces, the equilibrium distribution is spatially uniform, and the velocity distribution follows the Maxwell-Boltzmann form $f_{eq}(x,v) = n_0 \varphi_0(v)$,
 where $n_0=N/V$ and
\begin{equation}
    \label{M-distri.}
    \varphi_0(v) = m\lambda_T e^{-\frac{mv^2}{2k_B T}}
\end{equation}
with $\lambda_T\equiv h\left(2\pi mk_BT\right)^{-1/2}$ being thermal wavelength of the particles. Here, $k_{B}$ is Boltzmann constant, $T$ is the temperature of the surrounding environment and $V$ being the volume of the system. It is emphasized here that the derivation of the Maxwell distribution does not involve Newton’s second law, but instead assumes that the distributions of the individual velocity components are isotropic and independent.
 
 By comparison, we conclude that $B(x,v) = \text{const.} \equiv B > 0$, and $A(x,v) = A(v) \propto -v$. Setting $A(v) = -\gamma v$ and substituting this into (\ref{eq f_eq}), and comparing with (\ref{M-distri.}), we obtain 
\begin{equation}
B = \frac{\gamma}{m} k_{B} T.
\end{equation}
Thus, the evolution equation for the distribution due to collisions becomes 
\begin{equation}
 \label{eq FPE_of_velocity}
\frac{\partial f}{\partial t} = \gamma \frac{\partial}{\partial v} \left[ v f \right] + \frac{\gamma}{m k_{B} T} \frac{\partial^{2}}{\partial v^{2}} f\equiv I[f].
\end{equation}

%\subsection{The collisionless case and the complete evolution equation of distribution}

Secondly, we turn to the collisionless case. Consider an arbitrary volume $\omega$ in $\mu$-space (assumed to be sufficiently small for generality). The number of particles within this volume is given by $N_{V} = N \int_{\omega} f(x,v,t) d\omega$.
In the presence of collisions, particle velocities change discontinuously, leading to the non-conservation of particle number within the phase space volume $\omega$. However, in the absence of collisions, particle number is conserved in $\omega$, and the following holds 
\begin{equation}
N \int_{\omega} \frac{\partial f}{\partial t} d\omega = - \int_{S} \mathbf{J} \cdot d\mathbf{S},
\end{equation}
where $S$ is the 6-dimensional closed surface enclosing $\omega$, and $d\mathbf{S}$ is the outward-pointing surface element. The particle current $\mathbf{J}$ has two contributions  the position-space current $\mathbf{J}_{x}= N f(x,v,t) v \hat{x}$ and the velocity-space current $\mathbf{J}_{v}= N f(x,v,t) a \hat{v}$ with
%\begin{align}
%\mathbf{J}_{x} &= N f(x,v,t) v \hat{x}, \\
%\mathbf{J}_{v} &= N f(x,v,t) a \hat{v}.
%\end{align}
 $\hat{x}$ and $\hat{v}$ denoting the unit vectors in the position and velocity directions, respectively. By definition, $\dot{x} = v$, and the acceleration $a = \dot{v}$ (in the absence of collisions) is assumed to be independent of velocity, possibly a function of position, and is referred to as mechanical acceleration. Using the divergence theorem (Gauss's theorem), we obtain 
\begin{equation}
\nabla \cdot \mathbf{J} = N v \frac{\partial f}{\partial x} + N a \frac{\partial f}{\partial v}.
\end{equation}
Therefore, the conservation of particle number implies  
\begin{equation}
\frac{\partial f}{\partial t} + v \frac{\partial f}{\partial x} + a \frac{\partial f}{\partial v} = 0.
\end{equation}
This is the evolution equation for the distribution function in the absence of collisions. In three dimensions, this becomes 
\begin{equation}
\frac{\partial f}{\partial t} + \vec{v} \cdot \nabla_{r} f + \mathbf{a} \cdot \nabla_{v} f = 0,
\end{equation}
where $\mathbf{a}$ represents the acceleration of a single particle in the absence of collisions (mechanical acceleration).

Therefore, the complete evolution equation of the distribution function is governed by the Fokker-Planck equation 
\begin{equation}
\frac{\partial f}{\partial t} = -v \frac{\partial f}{\partial x} - a \frac{\partial f}{\partial v} + \gamma \frac{\partial}{\partial v} (v f) + \frac{\gamma k_BT}{m} \frac{\partial^{2} f}{\partial v^{2}}.\label{eq FPE2}
\end{equation}
Next, we could derive the equation of motion from the distribution. It can be shown from \eqref{eq FPE2} that
%To do this, we rewrite the Fokker-Planck equation as 
%\begin{equation}
%\frac{\partial f}{\partial t} = -\frac{\partial}{\partial x}(vf) - \frac{\partial}{\partial v}[(a - \gamma v) f] + \frac{\partial^{2}}{\partial v^{2}} \left( \frac{\gamma}{m \beta} f \right).
%\end{equation}
%Let $dx$ and $dv$ represent the total changes in position and velocity, respectively, over a time interval $dt$ (considering both external forces and environmental effects).
%\begin{equation}
%\label{langevin-step1}
%\overline{\Delta v} = -\gamma v \Delta t, \quad \overline{\Delta v^{2}} = 2 \frac{\gamma}{m} k_{B} T \Delta t,
%\end{equation}
 the mean and variance of the velocity changes over time interval $\Delta t$ are $\overline{\Delta v} = \left(a-\gamma v\right) \Delta t$ and $\overline{\Delta v^{2}} = 2m^{-1} \gamma k_{B} T \Delta t$, respectively. These results indicate that velocity fluctuations are present, leading to the equation of motion 
\begin{equation}
\label{Langevin-equation}
\frac{dv}{dt} = a - \gamma v + R(t), 
\end{equation}
where $R(t)$ is a stochastic term satisfying $\left\langle R(t) \right\rangle = 0$ and $\left\langle R(t) R(t') \right\rangle = 2m^{-1} \gamma k_{B} T\delta(t - t')$.

Equation \eqref{Langevin-equation}, known as the Langevin equation, is a stochastic equation that describes the motion of a particle interacting with its environment. In the case of a dilute gas system, if we consider one of the particles, then the surrounding particles serve as its "environment." This equation shows that the environment introduces fluctuations into the particle’s velocity. Since the timescale of collisions is much shorter than that of the particle’s motion, we can treat the environment statistically. In this context, the environment can be modeled as an ensemble, specifically a canonical ensemble in our case. The environment’s effect on the system is captured statistically. To be specific,  the $-\gamma v$ term, depending on the properties of the interested particle, can be interpreted as feedback from the particle's dynamics, and only negative feedback will drive the particle's velocity back to the equilibrium Maxwell-Boltzmann distribution. The environmental fluctuations account for the stochastic part $R(t)$ of the particle's velocity evolution equation.

%Eq.\eqref{Langevin-equation} is called Langevin equation, which is a stochastic equation describing the motion of a particle interacting with its environment.
%When considering a single-particle system, the surrounding particles act as its "environment." Equation (\ref{Langevin-equation}) shows that the environment introduces fluctuations into the particle's velocity. Since the timescale of collisions is much shorter than the timescale of the particle's motion, we can treat the environment statistically. In this context, the environment can be modeled as an ensemble (in our case, a canonical ensemble). 

We reiterate that the derivation in this section does not rely on Newton's second law; that is to say, the equations of motion can be constructed purely through statistical considerations.

\section{Statistical force and the Emergence of Newton's second law}
\label{Statistical force and its generalization}

In this section, we redefine the concept of "force" within the framework of a canonical ensemble and show that the Newton's second law emerges when the system reaches equilibrium, namely, the statistical force converges to the Newtonian force at equilibrium state. 

%\subsection{Emergence of Newton's second law}
According to Einstein, the force is defined as 
\begin{equation}
F_{e}(x,t) \stackrel{\Delta}{=} \frac{k_{B}T}{n(x,t)} \frac{\partial n(x,t)}{\partial x}. \label{eq def_of_F_e}
\end{equation}
Note that the collision process, which mainly influences the velocity of the particles, are much fast than other processes such as the motion induced by external field. As a result, the velocity of particles acts as a fast variable compared to their position. On the timescale of positional changes, the velocity can be approximated as remaining in a Maxwell distribution. Therefore, the evolution equation for slow variable $x$, i.e., the particle number density $n(x,t)$, can be derived by Zwanzig projection approach\cite{Zwanzig}, and the result is  
\begin{equation}
\label{FPE-n}
\frac{\partial n(x,t)}{\partial t} = \frac{1}{ m \beta \gamma} \frac{\partial}{\partial x} \left( -m \beta a + \frac{\partial}{\partial x} \right) n(x,t),
\end{equation}
where $\beta \equiv (k_{B}T)^{-1}$. The statistical force $F_e(x,t)$ is thus obtained by solving this equation. For physical consideration, one can set the reflection boundary condition for Eq.\eqref{FPE-n}.  In general, the solution also depends on the initial conditions. Nevertheless, no matter what form the initial condition is, the system would eventually evolve to a unique equilibrium state $n_{eq}$, which is  guaranteed by the properties of $I[f]$\cite{VanKampen}.

At equilibrium, i.e., when $\partial n/\partial t = 0$, the equilibrium condition formally yields the Newton's second law, $F_{e}^{eq}=ma$, where $F_{e}^{eq}$ is the equilibrium value of statistical force \ref{eq def_of_F_e}, namely,
%\begin{equation}
%m \beta a n_{eq} = \frac{\partial n_{eq}}{\partial x},
%\end{equation}
%or equivalently 
\begin{equation}
\label{Fe1}
F_{e}^{eq} = \frac{k_BT}{n_{eq} } \frac{\partial n_{eq}}{\partial x}.
\end{equation}
 This result forms the core of the paper: Newton's second law emerges naturally at equilibrium. However, when particles are out of equilibrium, the forces acting on them will deviate from this law. 

We can further derive the Boltzmann distribution
$n_{eq}(x) \sim \exp(-\beta W(x))$ from the Maxwell distribution without invoking Newtonian mechanics, and then establish the connection between statistical force and external field.

Firstly, the separation of timescales between velocity and position ensures that the position and velocity distributions are independent, with the velocity following Maxwell's distribution law, i.e., 
\begin{equation}
    \label{assump.1}
    N f_{\text{eq}}(x,v) = n_{\text{eq}}(x) \varphi_0(v),
\end{equation}

We now transition to the energy representation, where we consider the number of particles with energy \(\varepsilon = \varepsilon(x,v) = W(x) + mv^2/2\). From probability theory, it is straightforward to derive the corresponding result as follows
\begin{align}
  \label{N_epsilon}
    N(\varepsilon) = \int \delta \left( W(x) + \frac{1}{2}mv^2 - \varepsilon \right) \cdot n_{\text{eq}}(x) \varphi_0(v) \, dx \, dv \notag \\
= m\lambda_T \int e^{-\beta(\varepsilon - W(x))} n_{\text{eq}}(x) \frac{1}{\sqrt{2m(\varepsilon - W(x))}} \, dx.  
\end{align}

We assume that the external potential $W(x)$ is very small compared to the average thermal energy $k_B T$, i.e., $\sqrt{2m(\varepsilon-W)} \approx \sqrt{2m\varepsilon}$. In this case, $N(\varepsilon)$ is approximated as
\begin{equation}
\label{N_epsilon_result}
    N(\varepsilon) = m\lambda_T \frac{1}{\sqrt{2m\varepsilon}} \left\langle e^{\beta W(x)} \right\rangle e^{-\beta \varepsilon},
\end{equation}
where $\left\langle e^{\beta W(x)} \right\rangle \equiv \int e^{\beta W(x)} n_{\text{eq}}(x) \, dx$.

Next, on the other hand, by definition, we write $N(\varepsilon)$ as
\begin{equation}
    \label{N_epsilon_op2}
    N(\varepsilon) = \rho(\varepsilon) g(\varepsilon).
\end{equation}
Here $\rho(\varepsilon)$ is the single-particle distribution with respect to energy. The density of states, $g(\varepsilon)$, in the weak field approximation, can be expressed as:
\begin{equation}
    \label{g_free}
    g(\varepsilon) \approx g_{free}(\varepsilon) = V (2m\varepsilon)^{-1/2},
\end{equation}
where $g_{free}(\varepsilon)$ represents the density of states for free particles.

Finally, comparing Eq. \eqref{N_epsilon_result} and Eq. \eqref{N_epsilon_op2}, we find
\begin{equation}
    \label{rho}
    \rho(\varepsilon) =\rho_0 e^{-\beta \varepsilon}
\end{equation}
with $\rho_0\equiv V^{-1}\left\langle e^{\beta W(x)} \right\rangle m\lambda_T$. 
Substitute the energy density $\varepsilon = W(x) + mv^2/2$ into the above expression, we can eventually obtain Maxwell-Boltzmann distribution:
\begin{equation}
    \label{MB-distri}
    f_{\text{eq}}(x, v) = \rho_0 e^{-\beta \left( \frac{1}{2} mv^2 + W(x) \right)}.
\end{equation}

We emphasize that the above derivation does not rely on Newton's second law, and the derivation of the Maxwell velocity distribution also does not require Newtonian mechanics.

As a result, the equilibrium force can be further written as 
\begin{equation}
\label{Fe2}
F_{e}^{eq}(x) = -\frac{\partial W}{\partial x}.
\end{equation}
We refer to $F_{e}^{(eq)}(x)$ as the mechanical force to highlight that it depends solely on the potential $W(x)$ and not on collisions. The mechanical force $F_{e}^{(eq)}$ transforms the configuration space distribution from a uniform distribution to an exponential one, corresponding to the Boltzmann distribution.

\section{The statistical force in velocity space}

In this section, we introduce the concept of statistical force in velocity space, which characterizes the average effect of collisions on particle dynamics. This force plays a dissipative role, governing how collisions influence the overall motion of particles.

For simplicity, we consider the case without W(x), and the corresponding equation of motion is
\begin{align}
\frac{dv}{dt} &= -\gamma v + R(t), \label{eq langevin_eqn_no_V(x)} 
\end{align}
where $R(t)$ is a stochastic term satisfying $\left\langle R(t) \right\rangle = 0$ and $\left\langle R(t) R(t') \right\rangle = 2m^{-1} \gamma k_{B} T\delta(t - t')$.

As mentioned before, the collision processes mainly influence the velocity-part distribution of the particle, which have much smaller timescale than the observed macro-motion, i.e., they barely have impacts on the macro-motion. Actually, according to Eq.\eqref{eq langevin_eqn_no_V(x)}, they generally lead to dissipation of the mean velocity (as well as energy) of the particles.
 We call this part of motion, caused by collisions, as thermal motion. 

Analogous to the mechanical force \eqref{eq def_of_F_e}, we can define a generalized statistical force, named dissipative force, to quantify the mean effects of the collision processes as
\begin{equation}
F_{\eta} \stackrel{\Delta}{=} \gamma \frac{1}{\varphi(v,t)} \frac{\partial \varphi(v,t)}{\partial v}.
\end{equation}
Here $\varphi(v,t)$ is the velocity distribution of single particle. Without loss of generality, we consider the case with the absence of external forces, the evolution equation of $\varphi(v,t)$ is thus 
\begin{equation}
\frac{\partial \varphi(v,t)}{\partial t} = \gamma \frac{\partial}{\partial v} \left( v \varphi(v,t) \right) + \frac{\gamma}{m \beta} \frac{\partial^{2} \varphi(v,t)}{\partial v^{2}}.
\end{equation}
The solution is
\begin{equation}
\varphi(v,t) = \int dv_{0} P(v_{0} \vert v,t) \varphi_{0}(v_{0}),
\end{equation}
where the transition probability is given by 
\begin{equation}
P(v_{0} \vert v,t) = \sqrt{\frac{m \beta}{2\pi (1 - e^{-2\gamma t})}} \exp\left[-\frac{m \beta (v - v_{0} e^{-\gamma t})^{2}}{2(1 - e^{-2\gamma t})}\right].
\end{equation}
Assuming the initial distribution $\varphi(v,0)$ is Maxwell distribution, we obtain 
\begin{equation}
\varphi(v,t) = \sqrt{\frac{m \beta}{2\pi (1 - e^{-2\gamma t})}} \exp\left[-\frac{m \beta v^{2}}{2}\right].
\end{equation}
The relaxation time for collisions is $\gamma^{-1}$. For $t \gg \gamma^{-1}$, the velocity distribution evolves to the equilibrium Maxwell-Boltzmann distribution $\varphi_0(v)$.

Consequently, the dissipative force is $F_\eta=-m\gamma v$, which is proportional to velocity and always opposes the direction of motion. 

To analyze its effect in detail, note that the evolution equation for the velocity distribution, Eq. (\ref{eq FPE_of_velocity}), is a diffusion-type equation. The first term represents the drift flow driven by the dissipative force, and the second term represents diffusion. In the absence of the dissipative force, the velocity evolution would resemble free diffusion, similar to Brownian motion in position space. In this case, the velocity fluctuations $\left\langle v^{2} \right\rangle$ would grow with time, $\propto t$, and eventually, for $t \gg \gamma^{-1}$, the fluctuations would diverge, and the velocity distribution would become uniform. However, the presence of the dissipative term counteracts the diffusion, ensuring that the final steady-state distribution remains the Maxwell-Boltzmann distribution.

\section{Two illustrations of the statistical force}
\label{Two illustrations of the statistical force}

In this section, we give two examples to illustrate the statistical force and show that, before the system reaches equilibrium state, the fact that the statistical force $F_e(x,t)$ is not equal to the Newtionian force $ma$ actually indicate that there exists the net effect of the competition between energy and entropy.

\subsection{Statistical force in a harmonic potential}

As the first example, we consider the case where the external field is a harmonic potential. Let us consider a Brownian particle ensemble with an external potential $W(x) = m \omega^2 x^2/2$ and an environmental temperature $T$. The evolution equation of the particle density follows the Fokker-Planck equation 
\begin{equation}
    \label{harmonic-case-n}
    \frac{\partial n(x,t)}{\partial t} = \frac{1}{m\beta\gamma} \frac{\partial}{\partial x} \left( m\beta \omega^2 x + \frac{\partial}{\partial x} \right) n(x,t),
\end{equation}
with the initial distribution $n(x,0) = \delta(x)$ and reflection boundary condition in general. Here, the acceleration $a = -\omega^2 x$ is determined using Eqs. \eqref{Fe1} and \eqref{Fe2}. The analytical solution to the diffusion equation \eqref{harmonic-case-n} is easily obtained as 
\begin{equation}
    \label{solution-n-harmonic-case}
    n(x,t) = \sqrt{\frac{m \beta \omega^2}{2\pi\left(1 - e^{-\omega^2 t/\gamma}\right)}} \exp \left[-\frac{m \beta \omega^2 x^2}{2\left(1 - e^{-\omega^2 t/\gamma}\right)}\right].
\end{equation}
From the definition \eqref{eq def_of_F_e}, we obtain the statistical force induced by the inhomogeneity of the density distribution at time $t$ as 
\begin{equation}
    \label{statistical-force-harmonic}
    F_e(x,t) = \frac{ma}{1 - \mathrm{exp}(-\omega^2 t/\gamma)}.
\end{equation}
This indicates that, before the system reaches equilibrium, the average force acting on the particle is larger than the usual force $ma$ given by Newton's law.

\begin{figure}
  \centering
  \includegraphics[width = 8cm]{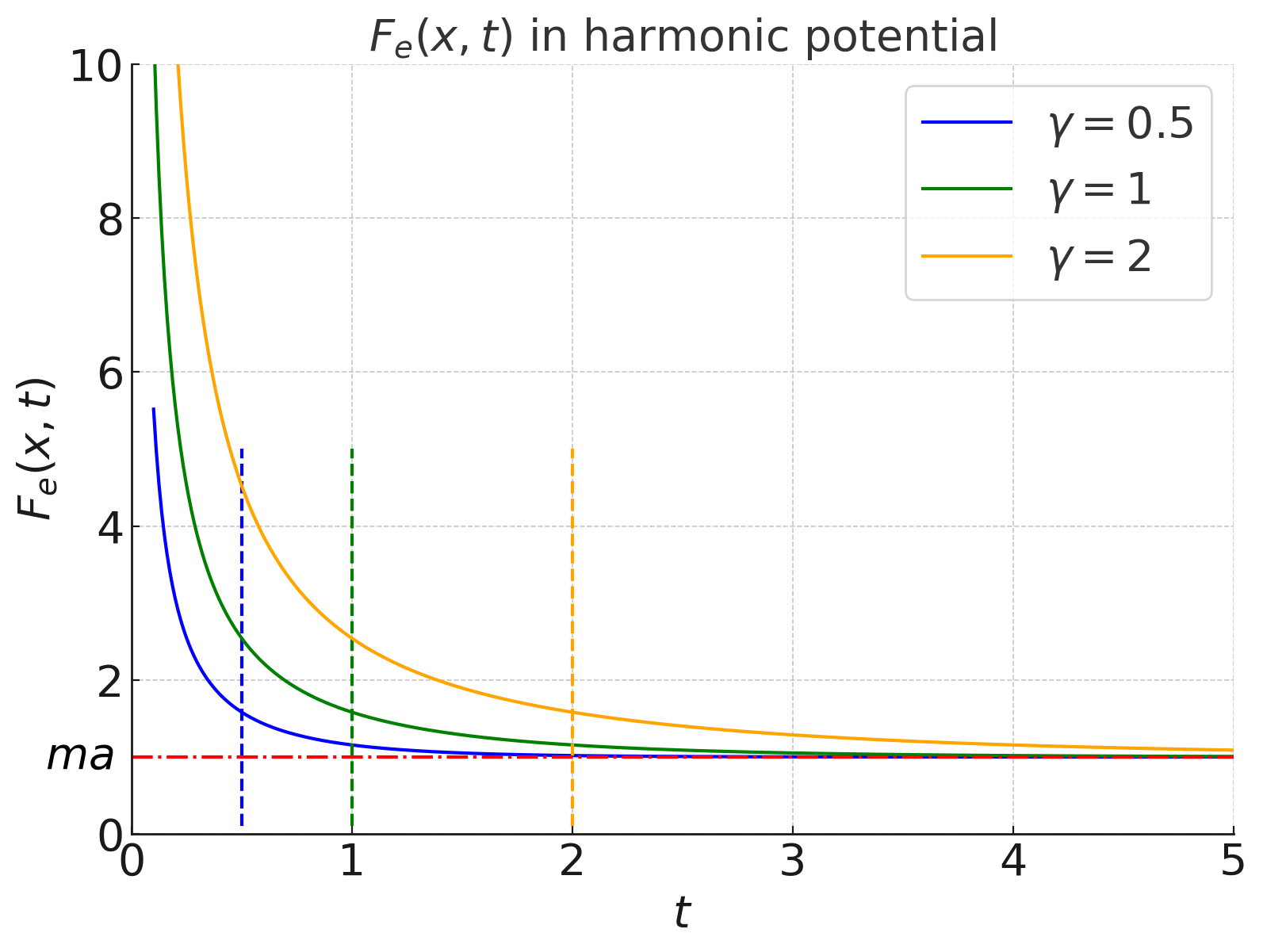}
  \caption{\label{Evolution of statistical force under harmonic potential} The statistical force in a harmonic potential. The blue, green, and orange solid lines correspond to $\gamma=0.5$, $\gamma=1$, and $\gamma=2$, respectively. The vertical dashed lines in corresponding colors represent the characteristic relaxation times $\gamma/\omega^2$ in different cases. The red dashed line represents the force $ma=-m\omega^2x$ predicted by Newton's law at equilibrium. As seen, all solid lines approach the horizontal line corresponding to $ma$ as $t$ increases, and eventually, at long times $t \gg \gamma/\omega^2$, $F^{(eq)}_e = ma$. For all three cases, we set $m = a = 1$.}
\end{figure}

Figure \ref{Evolution of statistical force under harmonic potential} shows the evolution of $F_e$ with time $t$ for different values of $\gamma$ under a harmonic potential, according to Eq.\eqref{statistical-force-harmonic}. In the long-time limit, i.e., when the particles are in equilibrium, Eq.\eqref{statistical-force-harmonic} naturally reduces to Newton's second law. However, before reaching equilibrium, the statistical force is larger than the usual Newtonian force, $ma$.

It is worth noting that the statistical force is actually proportional to the gradient of the entropy density $S(x,t) = -k_B \ln n$, representing the average entropy of particles at position $x$ at time $t$, i.e., $F_e = -T \partial_x S$, which is exactly the entropic force\cite{Verlinde,IgorMSokolov,RoosNico,RichardM.Neumann}. Moreover, Eqs.\eqref{Fe1} and \eqref{Fe2} tell us that $ma$ represents the direction of steepest descent of the energy density $\varepsilon(x) = k_BT/2 + W(x)$, which represents the average energy of particles at position $x$ at time $t$. Therefore, $F_e > ma$ indicates that, when particles are out of equilibrium, the net effect of the competition between energy and entropy drives the system in the direction of increasing free energy density
\begin{equation}
    F(x,t) = \varepsilon(x) - TS(x,t).
\end{equation}
This result may seem counterintuitive, but it does not imply a violation of the second law of thermodynamics. In fact, considering the total free energy $F(t) = \int F(x,t)n(x,t)\,dx$, the time derivative $\dot{F}$ of the free energy is
\begin{align}
    \frac{\partial F(t)}{\partial t} =& 
    \frac{n(x,t)}{m\gamma} \left(F(x,t) + k_B T\right) \left.\frac{\partial F(x,t)}{\partial x}\right|_{\mathrm{boundary}}\notag\\
    &- \frac{1}{m\gamma} \int n \left[\frac{\partial F(x,t)}{\partial x}\right]^2 dx,
\end{align}
where the first term on the right-hand side is a boundary term. Assuming that the particle density at the boundary can be neglected, this term vanishes. Therefore, the time derivative of the free energy is always negative until the system reaches equilibrium, where the derivative of the free energy becomes zero. This implies that the total free energy of the system decreases monotonically with time until it reaches its minimum at equilibrium.

\subsection{Statistical Force in Gravity field}

In this part, we consider the case in gravity field\cite{changpu-sun_chenliangqiti}. The external potential is $W(x) = mgx$ with $x>0$. The evolution equation of the particle density follows 
\begin{equation}
    \label{gravity-case-n}
    \frac{\partial n(x,t)}{\partial t} = \frac{g}{\gamma}\frac{\partial n}{\partial x}+\frac{1}{m\beta\gamma}\frac{\partial^2n}{\partial x^2}
\end{equation}
with the reflecting bottom 
\begin{equation}
    \label{reflecting-bottom}
    \frac{g}{\gamma}n(0,t)+\frac{1}{m\beta\gamma}\frac{\partial n(0,t)}{\partial x}=0.
\end{equation}
It is straightforward to derive the evolution equation of the corresponding statistical force $F_e(x,t)$ from Eq.\eqref{gravity-case-n}, namely,
\begin{equation}
    \label{Evo-F_e}
    \frac{\partial F_e(x,t)}{\partial t} = \frac{g}{\gamma}\frac{\partial F_e}{\partial x}+\frac{1}{m\beta\gamma}\frac{\partial^2F_e}{\partial x^2}+\frac{2}{m\gamma}F_e\frac{\partial F_e}{\partial x}
\end{equation}
with $F_e(0,t)=-mg$ and $F_e(x,t\rightarrow \infty)=-mg$. This is a nonlinear partial differential equation.The solution depends on the initial condition, and apparently, it has a trivial solution, namely, $F_e(x,t)=-mg$ for all $x$ and $t$ satisfying Eq.\eqref{Evo-F_e} and the boundary condition. In general, the statistical force is closely related to the initial density distribution, but no matter what form the initial condition is, the statistical force will eventually converge to the Newtonian force as Eq.\eqref{Fe1} shows.

To demonstrate, we choose initial condition to be $F_e(x,0)=-mg+\alpha x\mathrm{exp}(-x/\lambda)$ with $\alpha$ and $\lambda>0$ being constants. The corresponding initial density is $n(x, 0) = n_0 \exp\left[ - \beta m g x - \beta \alpha \exp(-x/\lambda) \left( \lambda x - \lambda^2 \right ) \right ]$. Here,  $n_0$ denotes the density at the ground level. A positive value of $ \alpha $ introduces a perturbation that increases the density near the ground, whereas a negative value produces the opposite effect, reducing the density in this region.

\begin{figure}
  \centering
  \includegraphics[width = 8cm]{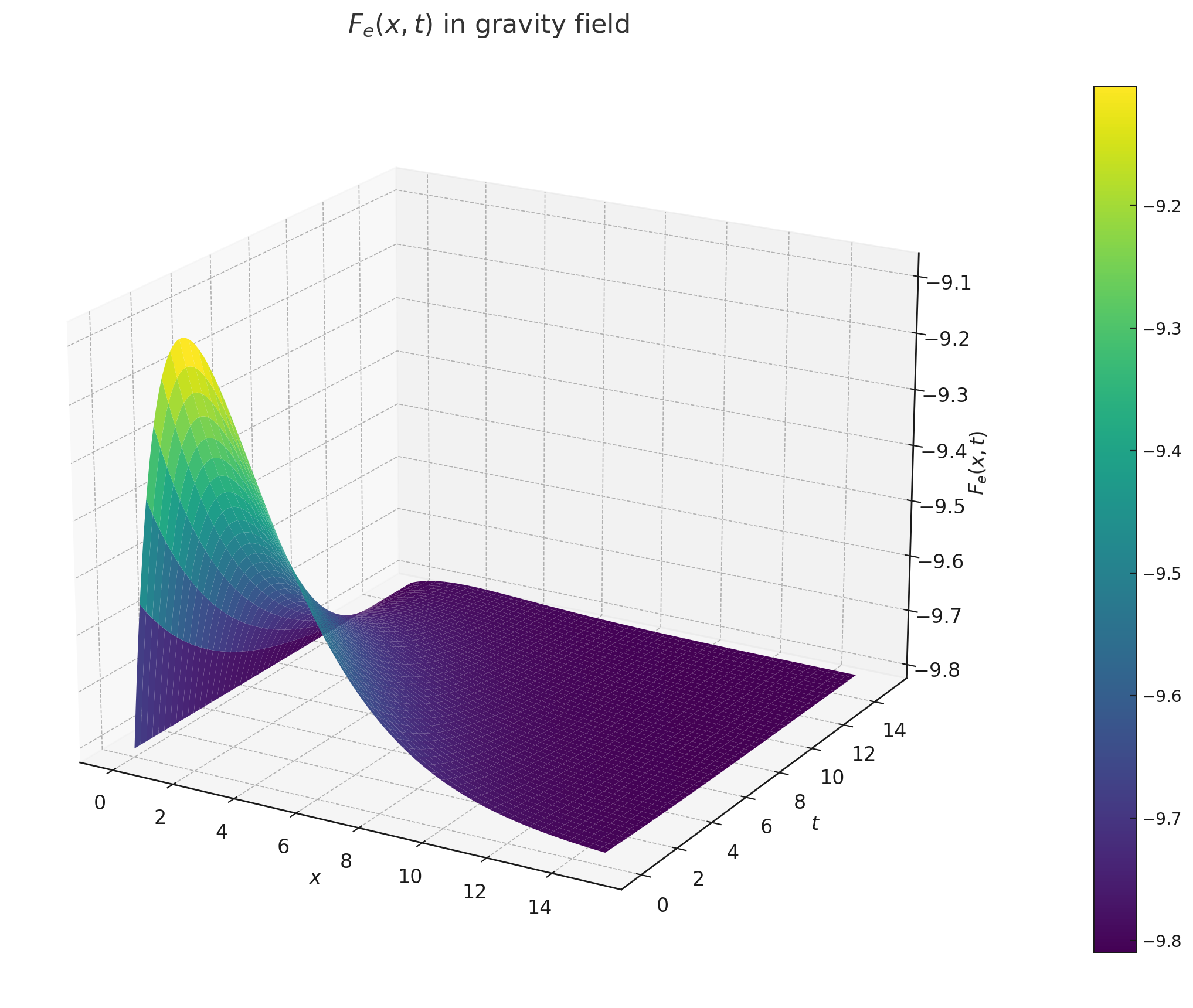}
  \caption{\label{eq.33}The statistical force in a gravity field. In this analysis, we fix the parameters as follows: \(\lambda = 2\), \(g = 9.8\), and \(\alpha = \beta = \gamma = m = 1\). For \(\alpha > 0\), the magnitude of the statistical force \(F_e(x, t)\) is initially less than the Newtonian force \(ma = -mg\) near \(x = 0\) and \(t = 0\). However, as \(x\) or \(t\) increase, \(F_e(x, t)\) asymptotically approaches the Newtonian value.}
\end{figure}
We find that the statistical force now behaves as 
\begin{equation}
    \label{F_e-special-gravity}
    F_e(x,t)=-mg+\alpha x\,\mathrm{exp}[-\frac{x}{\lambda}-\frac{t}{m\beta\gamma\lambda^2}].
\end{equation}
Apparently, when $x$ is small,i.e., not too high, the statistical force can exceed or be less than $-mg$, which depend on the sign of $\alpha$. Fig.\ref{eq.33} illustrates the case of $\alpha=1$, where the magnitude of the statistical force \(F_e(x, t)\) is initially less than the Newtonian force \(ma = -mg\) near \(x = 0\) and \(t = 0\) (For $\alpha < 0$, the magnitude of the statistical force is always less than the Newtonian force in the same region.). Nevertheless, in both cases, as $x$ or $t$ increase, $F_e(x, t)$ asymptotically approaches the Newtonian value.

\section{Conclusion}
\label{conclusion}

%In this work, we clarified the concept of statistical force as it arises from the inhomogeneity of particle distributions, using the Fokker-Planck framework. By redefining force in terms of distribution heterogeneity, we demonstrated that Newton’s second law naturally emerges as the system approaches equilibrium. Specifically, we derived the complete equation of motion for systems subject to external forces and environmental interactions, showing that the classical Langevin equation can be obtained directly by the evolution equation of the velocity distribution. In addition, we introduced the dissipative force characterized the impact of the environment and showed that it maintains the shape of the Maxwell distribution. Notably, the whole derivation does not make use of Newton's second law.

In this paper, we revisited the concept of statistical force as a manifestation of the inhomogeneity in particle distributions, utilizing the Fokker-Planck framework. By framing force in terms of distribution inhomogeneity, we demonstrated that Newton’s second law emerges naturally as the system approaches equilibrium. Specifically, we derived the complete equation of motion for systems influenced by external forces and environmental interactions, showing that the classical Langevin equation can be directly obtained from the evolution equation of the velocity distribution. Furthermore, we introduced the concept of dissipative force, which characterizes the environmental impact and preserves the shape of the Maxwell distribution. Importantly, our entire derivation proceeds without relying on Newton’s second law.

To illustrate this, we considered two specific cases: a harmonic potential and a gravitational field. For both, we derived the time evolution equation for the particle density and obtained an analytical expression for the statistical force at any given moment, assuming appropriate initial conditions. The statistical force, being proportional to the gradient of entropy density, deviates from the Newtonian force, which is related to the steepest descent of energy density. This deviation demonstrates that, in non-equilibrium regimes, net forces arise from competing effects between energy and entropy, with the latter depending on the initial state. However, regardless of the form of the initial conditions, the statistical force ultimately converges to the Newtonian force, as specified in Eq.\eqref{Fe1}.

In conclusion, this study provides a deeper understanding of the relationship between force and distribution inhomogeneity and reveals the important role of the entropic force, a form of statistical force, in the dynamics of non-equilibrium systems. The statistical force framework offers a new perspective on the interplay between thermodynamics and mechanics, with potential applications in systems governed by stochastic interactions, such as soft matter\cite{3ndLaw}, biological systems\cite{ao2005laws,wang2015landscape} (especially in neuroscience\cite{ramstead2023bayesian,friston2010free}), and active matter\cite{Vicsek,self-propelled,ivlev2015statistical,bechinger2016active,yuan2024quantum}.

\begin{acknowledgments}
This study is supported by NSF of China (Grants No.12088101), and NSAF (Grants No.U1930403 and No.U1930402).
\end{acknowledgments}

\bibliography{statistical_force_BibTeX_Export}%

% \newpage

\appendix
\begin{widetext}

\clearpage

% \bibliography{amm_corr}% Produces the bibliography via BibTeX.

\end{widetext}

\end{document}